# An Analysis of Data Driven, Decision-Making Capabilities of Managers in Banks

M. Shazmin Marikar and H.M.N. Dilum Bandara[1]

Department of Computer Science and Engineering

University of Moratuwa

Sri Lanka

{shazmin.16, dilumb}@cse.mrt.ac.lk

**ABSTRACT**

Organizations are adopting data analytics and Business Intelligence (BI) tools to gain insights from the past data, forecast future events, and to get timely and reliable information for decision making. While the tools are becoming mature, affordable, and more comfortable to use, it is also essential to understand whether the contemporary managers and leaders are ready for Data-Driven Decision Making (DDDM). We explore the extent the Decision Makers (DMs) utilize data and tools, as well as their ability to interpret various forms of outputs from tools and to apply those insights to gain competitive advantage. Our methodology was based on a qualitative survey, where we interviewed 12 DMs of six commercial banks in Sri Lanka at the branch, regional, and CTO, CIO, and Head of IT levels. We identified that on many occasions, DMs' intuition overrules the DDDM due to uncertainty, lack of trust, knowledge, and risk-taking. Moreover, it was identified that the quality of visualizations has a significant impact on the use of intuition by overruling DDDM. We further provide a set of recommendations on the adoption of BI tools and how to overcome the struggles faced while performing DDDM.

---

[1] Corresponding author

# 1 INTRODUCTION

Contemporary organizations are considering how to run smarter, be more agile, efficient, and competitive by using the right data to support efficient and effective decision making (Hall & Jia, 2015). In Data-Driven Decision Making (DDDM) we take data (structured or unstructured), analyzed them, and base a decision based on the analysis (Agrawal, 2014). DDDM is also known as "the practice of basing decisions on the analysis of data rather than purely on intuition" (Ransbotham, Kiron, & Prentice, 2015). Decision Makers (DMs) must make critical decisions based on the data to gain a competitive advantage in the dynamic business environment. Hence, DMs sought assistance from tools and technology.

Many tools and techniques support different steps of the decision-making process. Such tools are known as Business Intelligence (BI) tools, which is an umbrella term that refers to architectures, tools, databases, applications, and methodologies used to analyze data to support decisions of business managers. BI tools provide descriptive and predictive analytics of data with varying forms of visualizations to reduce the complexity of data analysis. Faster and better decision making, as well as insight creation, are among the top features where organizations intend to adopt BI tools (Hall & Jia, 2015). Consequently, organizations invest in expensive BI tools to support the DMs to process data and derive insights (Davenport, Harris, De Long, & Jacobson, 2001; Davenport, Harris, & Morison, 2010).

Banking has been a prolific industry for innovation where credit evaluation, branch performance, e-banking, and customer segmentation and retention are some of the best examples of the application of BI tools (Glaser & Strauss, 1967; Hall & Jia, 2015). Moreover, the rapid growth of data volume, structured and quality data, and availability of data have further influenced the baking industry to focus more on evidence-based decision making (Hensman & Sadler-Smith, 2011). However, to perform DDDM, DMs require a specific set of skills and capabilities such as data literacy, data interpretation, and understanding report and visualizations from the BI tools (Hensman & Sadler-Smith, 2011; Moro, Cortez, & Rita, 2015; Moro, Cortez, & Rita, 2014; Ransbotham, Kiron, & Prentice, 2015). While banks have invested heavily in BI tools and expect DMs to perform DDDM based on the outputs from BI tools, the effectiveness of such investment is being questioned as DMs seem to struggle while performing DDDM. Therefore, it is imperative to understand the underlying issues, as well as identify suitable remedies.

It is questioned that how leaders with limited analytical expertise can become adept consumers of analytics (Ransbotham et al., 2015). Another related question is whether the DMs know what they are looking in the data and whether the BI tool outputs are supportive in decision making. Therefore, how often data are transformed into valuable insights and business decisions is another question to answer. Many BI tool vendors and IT departments are also puzzled with questions such as why business leaders are not utilizing available tools for DDDM, and what should DMs be aware to make sure that their decisions are based on numbers, not intuition. Hence, it is imperative to understand the ability of the DMs to understand and interpret the data in a given business context, as well as the skill gap in DMs to make decisions based on evidence rather than pure intuition (Ransbotham et al., 2015; Yates & de Oliveira, 2016).

Several researchers have found that most organizations fail to use BI tools for DDDM. As seen in Fig. 1 several characteristics such as personal characteristics (Davenport, Dyché, & Schultz, 2013; Davenport et al., 2010), socio-demographic characteristics (Moro, Cortez, & Rita, 2014), DM's competencies(Yates & de Oliveira, 2016), and DM's data literacy (Dykes, 2017) influence effectiveness of DDDM. Ransbotham et al. (2015) highlighted the need for bridging the analytical gap among DMs to implement a successful BI solution. Moreover, researchers mention that DMs' personal characteristics and educational background have a more significant impact on practicing DDDM (Yates & de Oliveira, 2016). However, factors such as the discomfort in practicing DDDM, the impact of organizational culture on DDDM, the complexity of BI tools and their outputs, uncertainty due to performing Intuition-Based Decision Making (IBDM), DMs ability to interpret data, and data literacy have gained less attention by researchers. Moreover, existing studies have not focused on finding the struggles faced by the DMs while adopting and practicing DDDM, as well as suitable measures to address those pain points. Therefore, the problem to be addressed by this research can be stated as follows:

*What are the challenges faced by data-driven decision-makers and how to overcome them?*

We answer the above question through a qualitative survey and a detailed analysis of several cases of the use of BI tools. Our research focusses explicitly on the banking industry, where we interviewed 12 DMs of six commercial banks in Sri Lanka at the branch, regional, and CTO, CIO, and Head of IT levels. During the interviews, we focused on the extent the DMs utilize data and BI tools, as well as their ability to interpret various forms of outputs from BI tools and how they apply those insights to gain competitive advantage. It was identified that on many occasions, DMs' intuition overrules the DDDM due to uncertainty, lack of trust, knowledge, and risk-taking. Moreover, the quality of reports and visualizations from BI tools had a significant impact on the use of intuition by overruling DDDM. We further provide a set of recommendations on the adoption of BI tools and how to overcome the struggles faced when performing DDDM. Key recommendations include training and development on DDDM, gaining the trust of DMs on DDDM, and process changes to implement DDDM in the organization.

The rest of the paper is organized as follows. Section 2 elaborates on the research methodology and the prefigured factors. Section 3 provides an extensive analysis of the data collected via interviews and their analysis using the grounded theory approach. Recommendations for DMs, observations, and the key findings of the research are presented in Section 4, while concluding remarks are presented in Section 5.

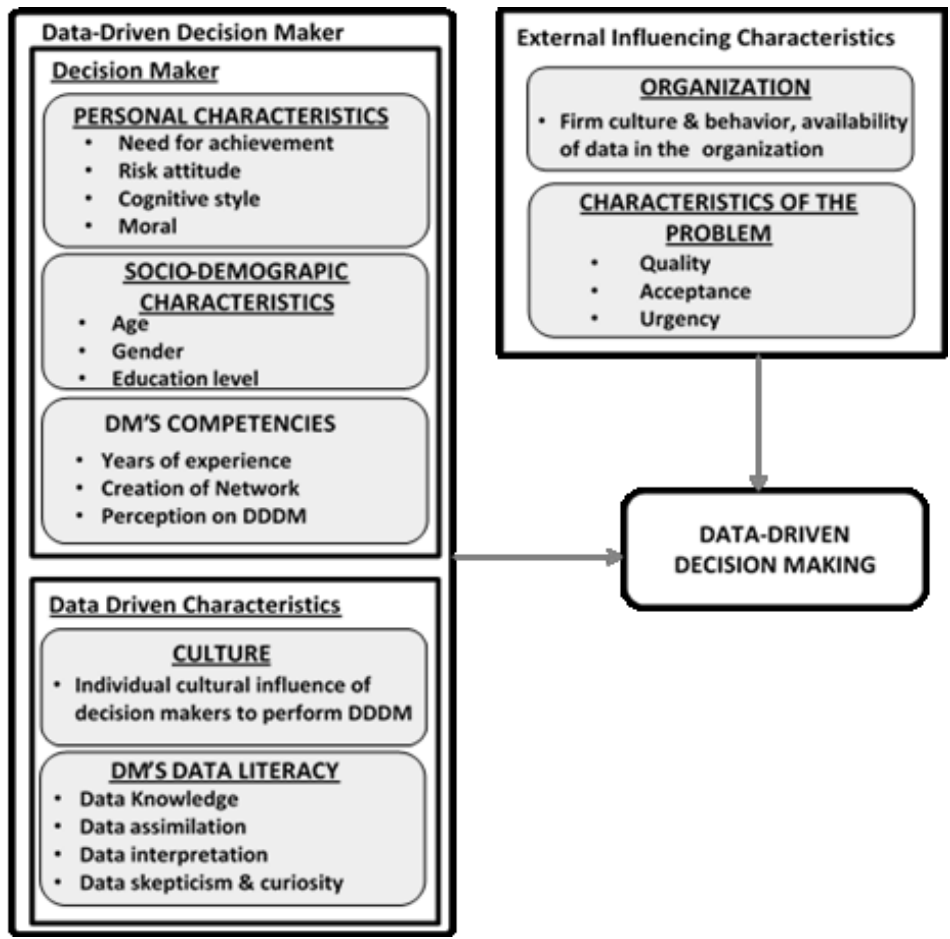

Figure 1: Characteristics influencing a DM to perform DDDM.

## 2 METHODOLOGY

Figure 2 presents the adopted research methodology. A literature review and three pilot interviews were initially conducted to identify the struggles DMs face and characteristics applicable in DDDM. Based on these findings, prefigured factors listed in Table 1 were derived. The dependent variable focuses on the DDDM ability of DMs. Independent variables focus on key factors uncovered through literature review and pilot interviews. Organizational culture was determined to be moderating the relationship between independent and dependent variables.

A set of interview questions was then derived based on the prefigured factors. Through theoretical sampling, a set of banks was chosen to conduct the interviews. Interview data were analyzed using the grounded theory approach while preparing a list of observations, findings, and recommendations. As there was no literature to identify a framework or a model to capture the DDDM related issues faced by the DMs, we aimed to develop a theory. Therefore, we used the Straussian grounded theory (Glaser & Strauss, 1967) to guide the research, which emphasizes the importance of coding data and identifying the critical findings through memo writing immediately after an interview.

Moreover, Straussian's version of grounded theory provides a more specific approach to identify the complex qualitative data into more meaningful coding. Furthermore, the Straussian-grounded theory is different from the classical Grounded theory, as it allows a literature review prior to the interviews. According to Strauss (Corbin & Strauss, 2008), literature helps theoretical sampling, concept development, and defining properties and dimensions. The objective of theoretical sampling is to generate a theory where the researcher collects, codes, and analyzes the data and decides what data to collect next and where to find them. After the initial interviews, data were coded and analyzed to guide the interview process further. While conducting the face-to-face interviews, we requested DMs to show sample reports and dashboards with the aim of observing the interpretive ability of the DM.

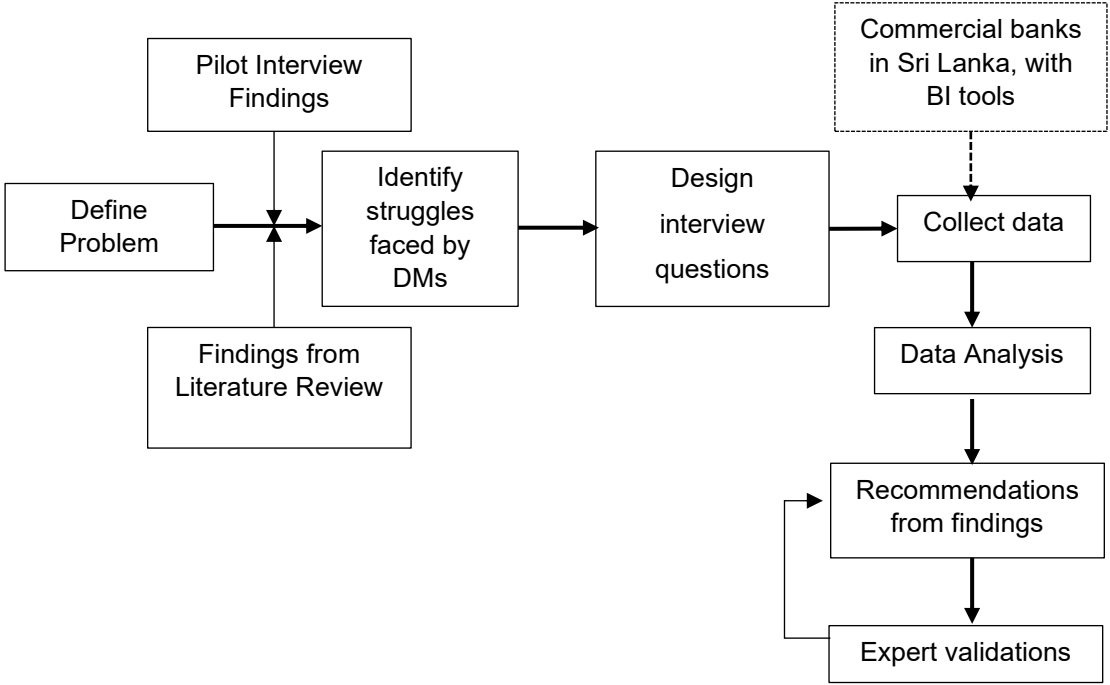

Figure 2: Research methodology.

Table 1: Dependent, moderating, and independent factors.

| Independent | Moderating | Dependent |
|---|---|---|
| Socio-demographic characteristics | Organizational culture | A DM's data-driven decision-making ability |
| Personal characteristics | | |
| DM's data literacy | | |
| DM's competencies | | |
| External characteristics influencing DMs | | |

We identified a set of banks, which use BI tools for decision making. As per the Central bank of Sri Lanka, there are 25 licensed commercial banks in Sri Lanka, where 18 of them are known to use BI tools (CBSL, 2017). We selected three local private banks, two state banks, and an international bank to ensure the study covers all the types of banks with BI implementations. To collect the opinions form different levels of decision-making hierarchy, we selected a branch-level manager, a regional-level manager, and CTO/CIO or the Head of IT. Details of the methodology, interviewees, and data analysis are available in (Marikar, 2018).

## 3. Data Analysis

### 3.1 Decision Makers' Competencies

A DM's competencies are classified into two aspects of DM's experience and DM's perception. Figures 3 and 4 show the dimensions related to DMs experience. As per the analysis of the interview data using the Straussian grounded theory, a relationship between the level of experience and IBDM was identified. As a DM's experience grows, he/she tends to make decisions based on intuition while overruling the DDDM process in the bank. For example, a regional manager of a semi-government bank stated that "While I gained experience, decision making is always intuition-based, but data is only used when I am unsure on a decision."

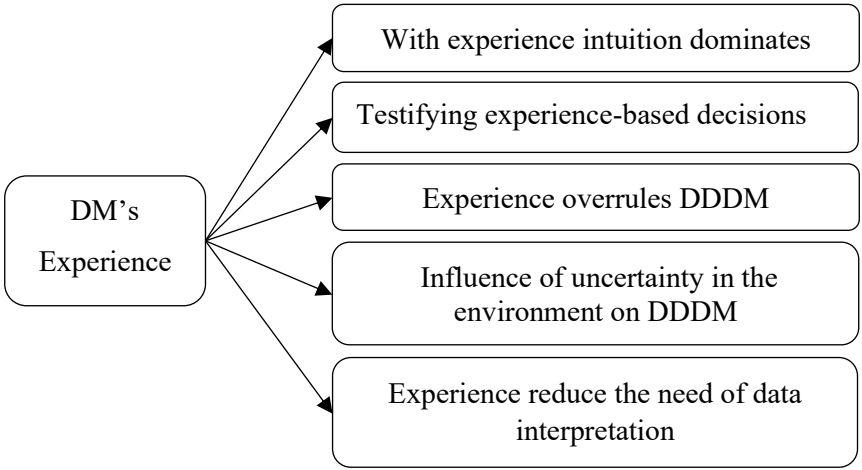

Figure 3: Dimensions related to DMs experience in DDDM.

Many managers make sure that the intuition-based decision is being supported and explained using the data, reports, and dashboards or tend to use the data to defend the intuition-based decision when the decision fails. For example, a manager of a private bank stated that "I do not use data to make decisions. But I use data to explain and check my decisions in the worst cases." This is counter-intuitive because, while we typically expect data to be used to derive a decision, most of the experienced managers seem to make intuitive decisions and then look for the data to support those decisions. This has been a challenge for CTO/CIO and Head of IT's in banks. For example, CTO from an international bank stated that "Managers do not use the BI tools as expected, but most of the time they cross-check the decision with data. But this is not our real expectation."

In a dynamic business environment, uncertainty is a dimension to be considered. Hence, many DMs are forced to consider such uncertainties. This is a root cause that DMs overrule the DDDM process and perform IBDM, which is due to the inability of BI tools to capture uncertainties in the business environment.

Lessons learned during a DM's service period motivated the use of IBDM over DDDM. Another important finding is that the experienced DMs have a low need for data interpretation, even if they currently use DDDM. This indicates that on many occasions, DMs intentionally overrules the DDDM based on their experience. An experienced DM from a private bank stated that "Lessons learn from the years of decision making cannot be replaced by a data graph."

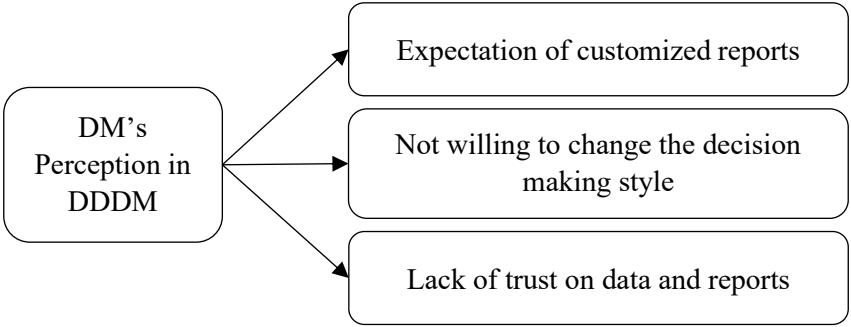

Figure 4: Dimensions related to DMs perception on DDDM.

### 3.2 Socio-Demographic Characteristics

As seen in Fig. 5 socio-demographic characteristics of a DM influence the DDDM. As per the analysis, knowledge, and exposure in the banking domain, experience in managerial positions, work exposure at different banks, and exposure on different geographies have an impact on the DDDM capabilities of a DM. Knowledge and exposure in the banking domain have a significant influence on the decision-making process, whereas DMs with vast knowledge and exposure tend to make more intuition-based decisions. For example, CTO from a private bank mentioned that "More than 50% of the decisions are made based on the gut feeling because the managers in higher positions are more experienced, and they do not see the real value of the data most of the times."

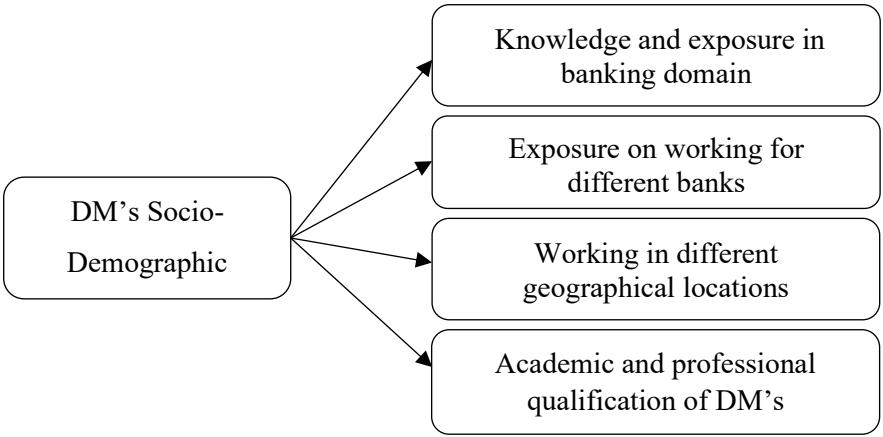

Figure 5: Dimensions related to DMs socio-demographic characteristics.

Based on open coding, it was identified that geographical location is key to choosing a decision-making approach. For example, most of the DMs away from the financial capital are not satisfied with the forecasting reports provided by the BI tools as the tools do not consider the geographical location of the branch as a factor in forecasting. In this case, a branch manager of an international bank pointed out that "The system provides good data. But since we are an international bank, the system is not specifically designed for our purpose like other local banks, so some reports are immaterial for us."

However, as a DM gains experience, recognizes patterns, and gets familiar with the system and tools, he/she tends to make more IBDM. Academic, professional, and domain-specific qualifications have a significant impact on a DM's approach. DMs with a STEM (Science, Technology, Engineering, and Mathematics) background are more influenced by their educational background to move towards DDDM.

**3.3 DM's Personal Characteristics**

A DM's personal character has both a positive and a negative influence on adopting and performing DDDM. As per the interview findings, influencing managers to follow a DDDM process is challenging. As seen in Fig. 6, personal characteristics of a DM such as risk-taking attitude, decision-making style (i.e., cognitive style), moral and individual perception, and willingness to change affect the decision-making style.

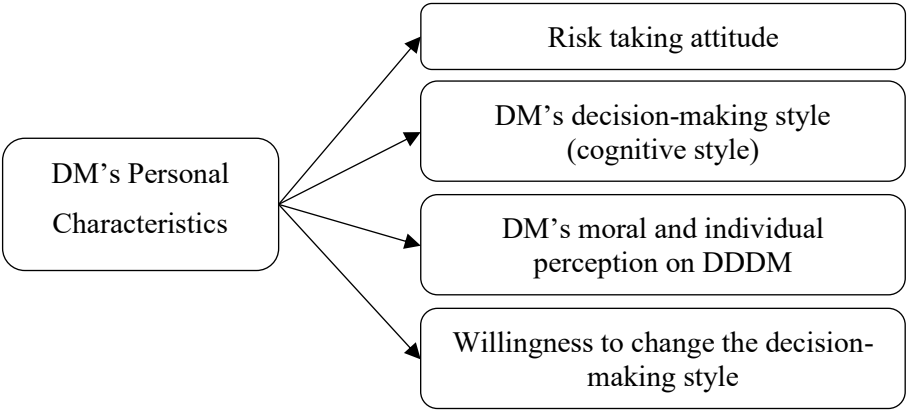

Figure 6: Dimensions related to DMs personal characteristics.

During the analysis, it was noticed that almost every DM is required to take risks to achieve given KPIs. However, risk-taking attitude among DMs differs, where some DMs are not willing to take risks due to bad experiences in the past. Alternatively, many inexperienced DMs are willing to take risks to achieve the KPIs, where these DMs expect support from analysis and forecasting reports/dashboards to mitigate risks and to ensure data support the decision. For example, a young manager of a private bank stated that "We need data to make a decision, and we have the ability to use data in decision making, but a certain amount of support is needed to understand data in some situations." Nevertheless, the underlying fact is that the DMs have the intention of using data and forecasting reports as a bailout from a worst-case scenario in case of a wrong decision.

A DM's cognitive style is a significant factor that influences him/her in adopting or practicing DDDM. According to the analysis, DMs find it challenging to move from a specific cognitive style to a different cognitive style. However, there are a few factors that influence a DM to change the cognitive style such as gut-feeling based decision making, the influence of the senior management, and the unique blend of DDDM and IBDM. As per the analysis, younger managers can be influenced more effectively to change the cognitive style, as younger DMs require more knowledge and support from both the data and higher management to make successful decisions.

Morals and the perception of individual DM also have an impact. While some of the DMs are willing to change the cognitive style, several underlying factors resist this change. These include lack of trust, uncertainty, and lack of experience and knowledge in DDDM. Experienced DMs are more comfortable in making decisions based on intuition or gut feeling, and when those decisions are successful, DMs develop an ego that negatively influences the willingness to change the cognitive style.

### 3.4 Data Literacy of Decision Makers

Every industry has its own set of unique terminology and data. It is required that DMs understand the data from the banking perspective. However, it is identified that most of the DMs do not have a clear idea or knowledge about the data available in the BI tools provided by the bank. For example, the Head of IT from a private bank stated that "Many managers struggle in matching data with business and making decisions, which is a major area to be focused." This is a significant drawback in enforcing DDDM, where the inability to understand the data in business terms, lack of basic statistical knowledge, and lack of knowledge on domain-related data are the critical struggles faced by the DMs.

Unfamiliarity with the reports and dashboards further influence IBDM. DMs with a statistical and STEM background find it easier to get familiar with the data compared to the DMs without such a background. For example, a branch manager mentioned that "It took me a considerable amount of time and a few years of experience to get used to data, by the way, I had to play a lot with data to get familiarized."

DMs seem to consider only the shape of the graphs and do not consider the units of measurements or magnitude. Alternatively, CTOs, CIOs, and head of ITs recommend training younger managers to help them understand the importance and the usefulness of the data assimilation, which could set the platform to improve and use more DDDM under any environment. During the interviews, a regional sales manager was requested to show a graph and interpret the graph. However, his interpretation was, "I only need to see the shape of the graph rest I can decide what to do with my experience."

Low-quality data visualization is another factor that requires attention from IT departments. Many DMs complain about the fidelity of the visualization, e.g., DMs need to identify the interest rate fluctuation.

As observed during the interviews, DMs' had difficulties in identifying the fluctuation clearly due to the unclear and low-fidelity visualization from the BI tools, which dissatisfied the DMs. This has been a critical reason for DMs to avoid practicing DDDM and perform IBDM even while making crucial decisions. Alternatively, low-quality visualizations and DMs' inability to interpret given reports and dashboards have created a need for self-service BI tools. A branch manager of a private bank and a regional manager of a state bank insisted that "I prefer customized reports, as well as to have my own reports which will ease my decision making instead of looking for data in different reports." However, high management has a different view on providing self-service BI tools for DMs, where they believe providing self-service BI tools will lead to more issues and more failures in DDDM. This is because such customized reports and graphs will not be tested for accuracy; hence, they might mislead DMs while making decisions.

Moreover, higher management has a concern about the accuracy of data used for such self-service reports and dashboards, as they seem to doubt some of the data stored in the systems and data warehouse. Alternatively, the reports in existing BI tools are well tested and known to be accurate. However, DMs argue that the reports are not specific to a region or a location where standard reports produced by the system do not provide the required information or data to perform DDDM. Whereas CTOs, CIOs, and Head of ITs counter-argue that the systems are providing the required reports, and it is the DMs who are not aware of the availability and the use of those reports. For example, a senior regional manager mentioned that "Many subordinates are no aware what report gives what information, and where to look for data this is a big issue I face."

Some DMs are highly dependent on specific reports and demand for modifications to the reports and dashboards to visualize the data in a more meaningful manner. The IT departments do not address such requirements; hence, remains a conflict between the IT departments and the DMs. Table 2 compares the views of DMs and CTOs, CIOs, and Head of IT's opinion of DMs' complains, requirements, and suggestions.

Banks encourage DMs to gain domain knowledge by gaining banking-related professional certifications. CTOs, CIOs, or Head of ITs suggest conducting training and hands-on sessions to transfer the data knowledge to bridge the gap in DMs.

As per the analysis, skepticism and curiosity were identified as significant factors of resistance against DDDM. Underlying reasons for skepticism and curiosity are the credibility and truthfulness. Almost all the interviewees highlighted that the credibility and truthfulness of the data, reports, and dashboards as major concerns while enforcing DDDM. The DMs are skeptical about the correctness of the reports provided; hens, tend to consider IBDM over DDDM. It was identified that many DMs tend to perform manual calculations to check the accuracy of reports and to verify them. Head of IT of a private bank mentioned that he experienced a situation where "I had discussions with many mangers. Due to bad experiences, they are not willing to use BI tools and reports, which is a perception that is hard to change." Moreover, DMs complained that the reports and dashboards consist of many assumptions when it comes to visualization, where they stated that such assumptions lead to inaccurate interpretation of the reports. Even though the assumptions are being communicated to the DMs, the understandability of the stated assumption is also low.

Table 2: Comparison of DMs and CTO's / CIO's or head of IT's opinions.

| DMs Requirement | Higher Management Opinion |
| --- | --- |
| Provide self- service BI tools for creating own reports. | Self-service BI can lead to data misjudgment & inaccurate reports & dashboards, which eventually lead to the failure in decision making. |
| Reports need to be customized for regions. | Reports can be flirted for specific regions, having specific region-wise customized reports will create overheads in systems. |
| Use only the basic reports and dashboards. | Other reports are requested to be self-learned; however, DMs have not considered self-learning. |
| System reports do not have the necessary information required for decision making in one place. | Systems reports have all the required information & data. Many DMs are not familiar with what is provided by the reports and where to find them on the system. |
| Reports need to be modified & updated based on the user needs | It is challenging to update a report only based on a specific person's requirement. It needs to be analyzed & redesigned to cater the new requirement, which takes a considerable amount of time. |
| Display data in tabular format under the graphs or dashboards. | This cannot be done with the limitations and visualization issues. There is no meaning in displaying data under a graph. |

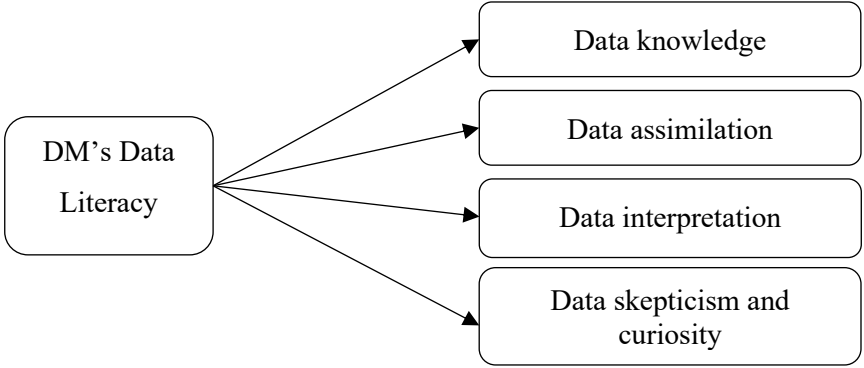

Figure 7: Dimensions related to a DM's data literacy.

Issues with data visualization, limitations of BI tools, and inability to generate customized reports are some of the critical issues identified during the study. During an interview one of the DM's pointed out that "The visualizations are not clear and identifying the difference between two data points is difficult to me, so this report is not useful at all." Moreover, another DM pointed out that "The graph's Y-axis is being scaled. However, in my case, the scaling is not useful, and it is useful only for the managers who handle a huge number of transactions."

These issues are related to technical and implementation limitation of BI tools; hence, they can be corrected relatively easily. When those issues are not addressed, DMs neglect the visualizations and make their own decisions based on intuition. As per the data analysis, it can be concluded that DMs data literacy is a significant factor that impacts adopting and practicing DDDM. Moreover, DMs' competencies, socio-demographic characteristics, and personal characteristics have a direct influence. During the study, several other factors such as organizational culture, competition among banks, networking among professionals, age, gender, and organizational behavior were identified to have a moderate impact on DDDM practices.

**3.5 Individual and Team Culture**

A DM's individualistic culture has a direct impact on adopting or practicing DDDM. Moreover, some banks have a team culture where a team of DMs performs DDDM, even though the individual DMs or their branches have diverse cultures. This seems to be motivated by the division of responsibility, where each DM's reputation will not be at stake in case the data-driven decision fails. Therefore, individual and team culture changes from person-to-person or team-to-team and has an impact on the adoption of DDDM. Moreover, different individuals and teams in the same banks tend to use different approaches.

During the study, it was further identified that when a bank has younger DMs who are willing to perform DDDM, and senior managers with the STEM and good statistical background, it creates a more conducive culture for adopting and performing DDDM. In this regard, a CIO of a private bank mentioned that "When we impose new technology like BI tools, most of the time the problem is with experienced old-school managers. But younger managers are more eager to adopt new technology."

**4. RECOMMENDATIONS**

To be successful in dynamic, uncertain, and competitive market, making strategic, operational, and branch-level decisions are crucial for the banks. While banks have adopted DDDM to gain competitive advantage, our findings indicate that the DMs in banks struggle to adopt and perform DDDM. This is a problem for the DMs as well as to the higher management who takes strategic initiatives and provide data and BI tools to support DDDM. Table 3 and Table 4 summarize the key observations and findings from the study, respectively, while key recommendations are listed in Table 5.

A DM's competencies can be categorized under the DM's experience and perception. Further, a DM's experience can be grouped into several dimensions, such as the domination of intuition, testifying the experience-based decisions using data, ignoring data and reports due to experience, and experience reducing the need for detailed data analysis. It was identified that as DMs gain more experience, the intention of practicing or adopting DDDM is low, where many experienced DMs resist the use of data, reports, and dashboards.

Even though intuition is dominating, due to the dynamic and uncertain business environment in Sri Lanka DMs are forced to consider DDDM, where DMs rely on a few reports and dashboards to identify the current situation. However, this could be considered as the basic level of DDDM. Because DMs are unable to extensively rely on such reports and dashboards, as they do not capture the uncertainty in the business environment. Thus, still, the IBDM is dominating in the Sri Lankan banking industry, especially among most of the well-experienced DMs regardless of the type of banks.

Testifying the accuracy, quality, and the acceptance of intuition-based decisions by referring to data is another practice of experienced DMs. This was performed by many DMs to ensure that the data will support their IBDM, as well as data as a backup to justify their decision for higher management in case a decision goes wrong. While the former can be considered acceptable, as the data is at least checked to ensure the accuracy of made decisions, the latter gives the perception that DDDM is not useful and when used could lead to failure.

Table 3: Key observations made during the study.

| **Uncertainty and Lack of Trust** |
|---|
| • The intention of getting support & backup from data to defend IBDM |
| • Testifying intuition-based decisions by conducting a cross-reference with data, reports, & dashboards |
| • Lack of data knowledge & trust in the existing system |
| **Need for Self-Service BI** |
| • Eager to use self-service BI tools for creating own reports |
| • Higher management is not in favor of providing self-service BI tools to DMs |
| • Not providing self-service BI tools lead to IBDM |
| • Concerns about accuracy & incorrect decisions |
| **DMs Personal and Working Environment** |
| • DMs away from the financial capital are not satisfied with the forecasting reports provided |
| • DMs blend IBDM & DDDM in many occasions |
| • DMs with a STEM & statistical background are more comfortable in using DDDM. DMs request for customized reports & data |
| **Risk Taking Attitude** |
| • DMs are not willing to take the risk due to bad experiences in the past |
| • DMs with less experience are more eager to take risks |
| • DMs have the intention of getting support and backup from data to defend IBDM when they fail. |

During the study, we identified a set of DMs who are eager to use DDDM. These DMs have unique characteristics such as willingness to take risks, support, try to achieve a competitive advantage from data, and elaborate decisions to higher management by providing evidence form data. However, it was further identified that the ego developed when the IBDM results in success and confidence in making decisions based only on intuition than DDDM as key reasons where IBDM overrules the DDDM. Hence, the challenge for banks is to gain trust and increase the confidence of DMs' on using DDDM as a day-to-day practice.

Apart from the experience of a DM, the perception of DMs on DDDM is also another aspect. DMs perception includes the lack of trust and expectation of customized reports and data. Most of the DMs tend to perform manual calculations on the values/figures provided by the BI tools to ensure they are getting the correct values. This exemplifies the lack of trust in the data, reports, and dashboards or even the developers that developed them. After performing manual calculations for a while, DMs gain the trust on data and reports, yet DMs do not have the confidence to drive the decision based on the data. In this case, the recommendation for the banks would be to ensure that the DMs gain the confidence that the data could drive them towards successful decision making.

Socio-demographic characteristics of a DM is a combination of a distinct set of dimensions such as knowledge and exposure in the banking domain, academic and professional qualifications of DMs, and exposure of working for different banks and different geographical locations. During the study, it was identified that mapping data to a business problem is a significant struggle DMs face when considering DDDM. However, having in-depth knowledge and exposure to distinct levels of banking domain can provide a strong background for DMs to map data to the business problems. Alternatively, academic, professional, and domain-related qualifications have a positive impact on adopting and practicing DDDM. Moreover, DMs with a STEM and statistical background are more comfortable in DDDM comparative to other DMs, where they practice and adopt DDDM regularly.

The geographical location of a branch has a significant impact on whether the DMs practice DDDM. As per the analysis, DMs away from the commercial capital tends to use more DDDM with the aim of achieving KPIs. However, the DMs in the commercial capital regularly achieve KPIs; hence, the need for data is low. This indicates that when DMs exhaust all other options to improve their KPIs, they rely on data to gain insights and make decisions. Moreover, branches in commercial capital seem to have lower KPIs, as DMs are comfortable in relying only on intuition to reach their KPIs. This is an indication that bank's higher management has not investigated data to identify who has achieved the KPIs and the need to raise the bar. Moreover, the inability of BI tools to present results based on user roles and branch-specific circumstances deter the use of DDDM. Users have lost the trust in BI tools, as they are not adaptive to specific needs, as well as the inconsideration of higher management on user and branch-specific challenges.

In addition to DMs competencies and socio-demographic factors, DMs personal characteristics have a considerable influence on adopting and practicing DDDM. DMs personal characteristics are categorized as risk-taking attitude, decision-making style (cognitive style), DMs moral and individual perception on DDDM, and willingness to change the cognitive style. Most of the experienced DMs are not willing to take risks as they had bad experiences in the past. Nevertheless, many inexperienced DMs are willing to take risks with the aim of achieving KPIs where these DMs expect support from data analysis and forecasting reports to mitigate risk and to ensure the decision is supported by data analysis. While each DM has a unique style of decision making, banks expect a DM to change his/her cognitive style to adopt and practice DDDM while overcoming uncertainty, as well as lack of trust, knowledge, and experience.

Data literacy of DMs is considered a significant factor in this study, whereas it was observed as a significant struggle faced by the DMs. Data literacy is classified as data knowledge, data assimilation, data interpretation, and skepticism and curiosity. DMs experience a knowledge gap due to the lack of ability to understand the data in the business context, not having basic knowledge of statistics, and lack of understanding of domain-specific data. Data assimilation (i.e., familiarizing with the data presented to the DM) is a key attribute that a DM requires to perform effective DDDM. Moreover, most DMs are not aware of the availability of data, what data is available in which report, and how to find them.

Table 4: Summary of findings from the study.

| |
|---|
| **Data Visualization Issues** |
| • Lack of fidelity, quality, & clarity of visualization have a negative influence on DMs, where they resist using reports and dashboards for decision making. This leads DMs to request self-service BI tools |
| • Some graphs & charts are not used at all due to poor visualization. |
| **Personal Characteristic of DMs** |
| • DMs with STEM & statistical education background are more comfortable in adoption & practicing DDDM comparative to other DMs |
| • Younger DMs can be influenced to adopt & practice DDDM compared to more experienced DMs |
| • DMs intentionally overrule DDDM & make decisions with intuition due to experience gained while practicing DDDM |
| **Domain Knowledge Related Issues, Uncertainty, and Lack of Trust in DMs** |
| • Lack of familiarizing of the system provided reports, dashboards, & data lead to IBDM |
| • Self-service BI tools will lead to more issues & errors, as reports/graphs will not be tested for accuracy, which might mislead DMs in decision making |
| • There is a lack of trust about the accuracy of data in the system/tools among all levels of DMs |
| • Uncertainty, lack of trust, knowledge, and experience in DDDM create resistance on DMs willingness to change their cognitive style |
| • Lack of ability in understanding the data in the business perspective, domain-related data, & not having a basic knowledge of statistics are underlying factors for data knowledge gap among DMs |
| **Risk Taking Attitude of DMs** |
| • Experienced DMs are not willing to take risks |
| • Younger DMs with less experience are willing to take risks as they expect data to support & mitigate risk |

These lead DMs to request new reports and to use self-service BI tools for creating their own reports. However, according to Head of ITs, CTOs, and CIOs of banks, this is not a positive move, as they argue that the use of self-service BI tools has the potential risks of making false conclusions and wrong decisions due to the inaccuracy and errors in data and custom reports. Alternatively, DMs argue that the existing reports do not cater to their data needs, and customized reports could give them a unique advantage to achieve their KPIs. Some DMs complain that not providing self-service BI tools leads to IBDM, as DMs are unable to find data to support or oppose their innovative ideas. The first problem of lack of awareness could be addressed through better user training on BI tools and what resources are available. However, this also indicates that higher management does not trust the data on the system;

hence, fear custom reports could be wrong as the underlying data may be inaccurate. Moreover, they seem to not trust their subordinates' competencies to use self-service BI tools. Hence, they prefer to restrict the options and double-check every such option, report, and dashboard produced by the BI tools. This also indicates that BI tools are not mature enough to ensure that only the appropriate graphs and reports can be generated given the role of the DM and business domain, e.g., given some numeric values, BI tools may easily generate a wrong graph.

Many DMs were only concerned about the shape of the provided graphs than the numerical values they reflect. The reason for this is that the DMs are used for these graphs over time, and metrics remain the same. Due to the inferior quality of visualizations, DMs are not using some of the reports and rely on making their own calculations. DMs favor having data in tabular format than graphs or charts and even prefer to see tabular data below the visualization. This is due to the fact that they are trained for spreadsheets than visual analytics and interpretations. Moreover, DMs personal characteristics and socio-demographic characteristics have a direct relationship to their ability and intention to use DDDM.

Table 5: Recommendations derived from the study.

| **Training and Development** |
| --- |
| - Make DMs aware of what tools, reports, and data are already available |
| - Provide training & conduct practical sessions with the aim of reducing data literacy gap |
| - Focus on training younger DMs towards DDDM compared to well-experienced DMs |
| - Train DMs on basic statistics & how to map data to business domain |
| - Encourage DMs to gain domain knowledge through professional qualifications |
| **Trust and confidence of DDDM** |
| - Gain DMs trust by presenting accurate data, reports, & graphs |
| - Improve quality & clarify of visualizations |
| - Provide self-service BI tools & some custom reports as they may have innovative ideas that they are looking for data to support |
| - Improve the confidence of DMs that DDDM can lead to success |
| - Provide support for DMs to make decisions based on data & provide them with a platform to testify DDDM against the intuition |
| **Process Changes** |
| - Use well-experienced DMs who perform DDDM as change agents to encourage DDDM |
| - Recognize successful cases of DDDM |
| - Encourage risk-taking based on DDDM |
| - Revisit KPIs to see whether they are pushing DMs to look beyond trivial |

Based on the observations, findings, and recommendations, we propose the framework to develop DDDM ability among DMs, influence practicing DDDM, and to provide a better understanding to organizations about the benefit of DDDM. The proposed framework shown in Fig. 8 consists of three maturity levels related to adoption, practicing, and influencing DM towards DDDM. The first vertical focuses on gaining the trust of DMs on DDDM, providing them the required knowledge to perform DDDM, as well as making DMs comfortable with data to perform DDDM over IBDM, and making sure all the required resources, BI tools, and process steps are implemented in banks to use data as a source for decision making. The practice phase is an icebreaker for DMs, where experienced DMs who are willing to change the DM approach, as well as DMs with a STEM or statistical background, could influence the DMs who are resisting to practice DDDM. This could provide the required level of influence and confidence from the fellow employees to move away from the comfort zone of IBDM, and to make sure they are aware of DDDM. The maturity phase of the framework is about the continuous practice of DDDM by the DMs, where successful DMs are recognized and rewarded, while unsuccessful practices are continuously revised or eliminated without being judgmental on DMs. Moreover, KPIs need to be revisited and adjusted accordingly as DDDM becomes a practice than an exception.

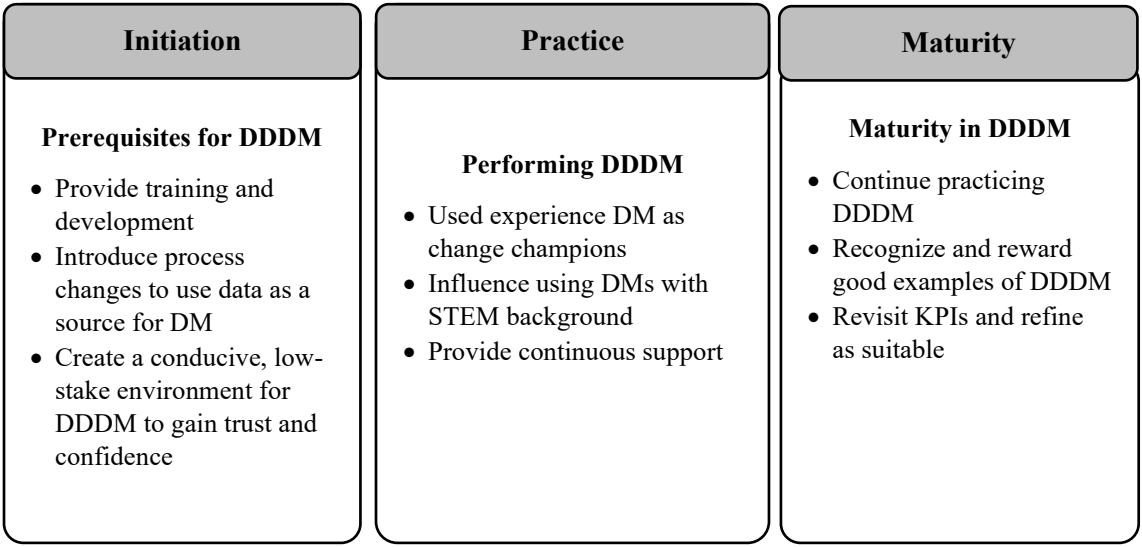

Figure 8: The proposed framework to adopt, practice, and influence DDDM.

## 5. Summary

Given the high competition, lower margins, as well as dynamic and uncertain business environment, banks have identified the importance of using data for decision making. While banks have invested heavily in BI tools and made data available to the DMs, our study found that DMs' competencies, socio-demographic characteristics, personal characteristics, and data literacy have limited the impact of adopting and practicing DDDM. Moreover, external factors such as organizational behavior and culture have a moderate impact on the DMs adopting and practicing DDDM. Use of data to support intuition-based decisions, lack of trust and uncertainty on DDDM, differences between younger and experienced DMs' risk-taking attitude, educational background (especially STEM education), need of self-service BI tools, and conflict among DMs and higher management are the unique findings of the study compared to the related work. While we derived the recommendations based on related work and interview findings, in the future, we plan to explore the effectiveness of the proposed recommendations in practice. Moreover, another related area of future work is to develop a scorecard to measure the DMs data literacy, as it is essential to recommend a suitable professional development plan designed specifically for each DM.